\documentclass[prl,aps,twocolumn,superscriptaddress,showpacs]{revtex4}

\usepackage{graphicx}

\usepackage{amsmath}

\usepackage{amssymb}

\begin{document}

\title{Complex shape evolution of electromigration-driven 
single-layer islands}

\author{Philipp Kuhn}

\affiliation{Fachbereich Physik, Universit\"at Duisburg-Essen,
45117 Essen, Germany}

\affiliation{Institut f\"ur Theoretische Physik,
Universit\"at zu K\"oln,
Z\"ulpicher Strasse 77,
50937 K\"oln,
Germany}

\author{Joachim Krug}

\email[]{krug@thp.uni-koeln.de}

\affiliation{Institut f\"ur Theoretische Physik,
Universit\"at zu K\"oln,
Z\"ulpicher Strasse 77,
50937 K\"oln,
Germany}

\author{Frank Hausser}

 \affiliation{Crystal Growth group, research center caesar,
    Ludwig-Erhard-Allee 2,
    53175 Bonn, Germany}

\author{Axel Voigt}

 \affiliation{Crystal Growth group, research center caesar,
    Ludwig-Erhard-Allee 2,
    53175 Bonn, Germany}
\date{\today}

\begin{abstract}

The shape evolution of two-dimensional islands through periphery
diffusion biased by an electromigration force is studied numerically using a continuum
approach. We show that the introduction of crystal anisotropy in the 
mobility of edge atoms induces a rich variety of migration modes,
which include oscillatory and irregular behavior. A phase diagram in
the plane of anisotropy strength and island size is constructed.
The oscillatory motion can be understood in terms of stable
facets which develop on one side of the island and which the island
then slides past. The facet orientations are determined analytically.

\end{abstract}


\maketitle

The manipulation of nanostructures by macroscopic forces is likely
to become a key ingredient in many nanotechnology applications.
Understanding the influence of external fields on the 
shape evolution of nanoscale surface features
is therefore of considerable importance. As a first step
in this direction we analyze here the effects of an
electric current on single-layer islands on a
crystalline surface.
The islands evolve under surface electromigration,
the directed motion of adsorbed 
atoms due to the slight force transmitted by collisions
with the conduction electrons in the sample \cite{Sorbello98}. 

Electromigration along interfaces and grain boundaries is 
the most persistent and menacing reliability problem in 
integrated circuit technology \cite{Suo03,Tu03}.
Correspondingly, much work has
been devoted to electromigration-induced void formation
and breakdown in metallic conductor lines \cite{Tu03},
and the capacity for quantitative numerical modeling has been
demonstrated at least for simple void geometries \cite{Gungor99,Schimschak00}. 
A major obstacle to achieving predictive power in such studies,
however, is the  
insufficient control over the complex internal
structure of the polycrystalline samples. Hence an important
motivation for investigating  
electromigration-induced effects on simple,
well-controlled nanoscale morphologies,
such as step patterns on vicinal  
surfaces \cite{Yagi01}
and single layer islands \cite{Metois}, 
is to bridge the gap between
the microscopic mechanisms of electromigration
and their consequences on technologically relevant
length and time scales.

Electromigration of islands has
been modeled previously
using Monte Carlo simulations 
\cite{Mehl00} and continuum theory \cite{PierreLouis00}.   
The continuum approach to island shape evolution, which treats
the island edge as a smooth
curve, has been successfully applied to a range of 
problems including the diffusion \cite{Khare95}
and sintering \cite{Liu02} of islands, 
and the pinch-off of vacancy clusters \cite{Pai01}.
Here we focus on the regime of periphery diffusion (PD),
where the dominant kinetic process is the migration of atoms
along the island boundary. The shape then 
follows a local evolution law, without coupling to the
adatom concentration on the surrounding terrace. 

We extend
the model of \cite{PierreLouis00} by including
crystal anisotropy in the adatom mobility. 
It was observed recently in the context of step flow
growth \cite{Danker03} that crystalline anisotropy
can change the behavior of step patterns in 
a \textit{qualitative} way.
In the present case, it leads to the unfolding of a remarkable richness
of dynamic phenomena: In addition to the scenarios
of steady motion and island breakup \cite{PierreLouis00,Schimschak00}
observed in previous work, we find spontaneous
symmetry breaking, oscillatory
shape evolution, and complex migration trajectories where different
modes of motion alternate in a periodic or irregular fashion. 
This highlights the importance of properly including anisotropy
in the modeling of boundary evolutions.   
Oscillatory shape dynamics has been seen previously in a numerical
study of void electromigration \cite{Gungor00}, and transitions
to quasiperiodicity and chaos are known to occur in  directional solidification
\cite{Kassner91,Valance92}. To the best of our knowledge, however,
our work provides the first example of complex shape evolution for
a \textit{closed} contour subject to purely \textit{local} dynamics. 

In the PD regime, the normal velocity $v_n$ of the island
boundary satisfies the continuity equation
\begin{equation}
\label{cont}
v_n + \frac{\partial}{\partial s} \Omega \sigma \left[
- \frac{\partial}{\partial s} (\Omega \tilde \gamma \kappa) + q^\ast E_t 
\right] = 0.
\end{equation}
Here $s$ denotes the arclength along the island edge, and 
$\Omega$ the atomic area.  The square bracket
multiplied by the edge atom mobility $\sigma$ is the total
mass current along the boundary, which is driven by the tangential
derivative of the chemical potential $\Delta \mu = \Omega \tilde \gamma
\kappa$ and the electromigration force $q^\ast E_t$;
$\tilde \gamma$ is the edge stiffness, $\kappa$ the local curvature,
$q^\ast$ the effective charge of an edge atom, and $E_t$ the tangential     
component of the local electric field. 
The crystal anisotropy of the surface enters through the dependence
of $\tilde \gamma$ \cite{Dieluweit03} and $\sigma$ \cite{Liu02,Giesen04} 
on the orientation angle $\theta$ of the island edge.  

For atomic layer height islands on the surface
of a thick sample, the island boundary has a negligible effect
on the electric field; this is in contrast to the
modeling of voids in conductor lines,
where the coupling of the void shape to the
electric field leads to a manifestly nonlocal
problem \cite{Gungor99,Schimschak00,Gungor00}. 
Here we can take the field to be of constant
strength $E_0$
and aligned along the $x$-axis. Letting $\theta$ denote the angle
between the normal of the island edge and the $y$-axis
(counted positive in the clockwise direction), this implies
$E_t = E_0 \cos(\theta)$. 

Together with the specification of 
$\tilde \gamma$ and $\sigma$, to be addressed below, this completes
the definition of the local boundary evolution (\ref{cont}).    
Comparing the two terms inside the square brackets, we 
extract the characteristic length scale \cite{PierreLouis00,Schimschak00,Suo94}
$l_E = \sqrt{\Omega \tilde \gamma /\vert  q^\ast E_0 \vert}$, which gauges
the relative importance of capillary and electromigration forces; electromigration
dominates on scales large compared to $l_E$.
Below all lengths are reported in units of $l_E$.

The isotropic version of (\ref{cont}), with $\tilde \gamma, \sigma = \mathrm{const.}$,
has been studied previously by Suo and collaborators \cite{Suo94,Wang94,Suo96}.
A circular island moving at constant velocity is stable for (dimensionless)
radii $R < R_c \approx 3.26$ \cite{Suo96}. Beyond the instability a bifurcation to two 
branches of non-circular stationary solutions occurs \cite{Wang94}. 
Numerical integration of the time-dependent problem \cite{Kuhn04}
shows that only one of the branches, corresponding to islands elongated
in the field direction, is realized. At large radii island breakup occurs, 
mediated by the outgrowth of a finger of the kind found in \cite{Suo94}. 

\begin{figure}
\centerline{\includegraphics[scale=0.25]{%
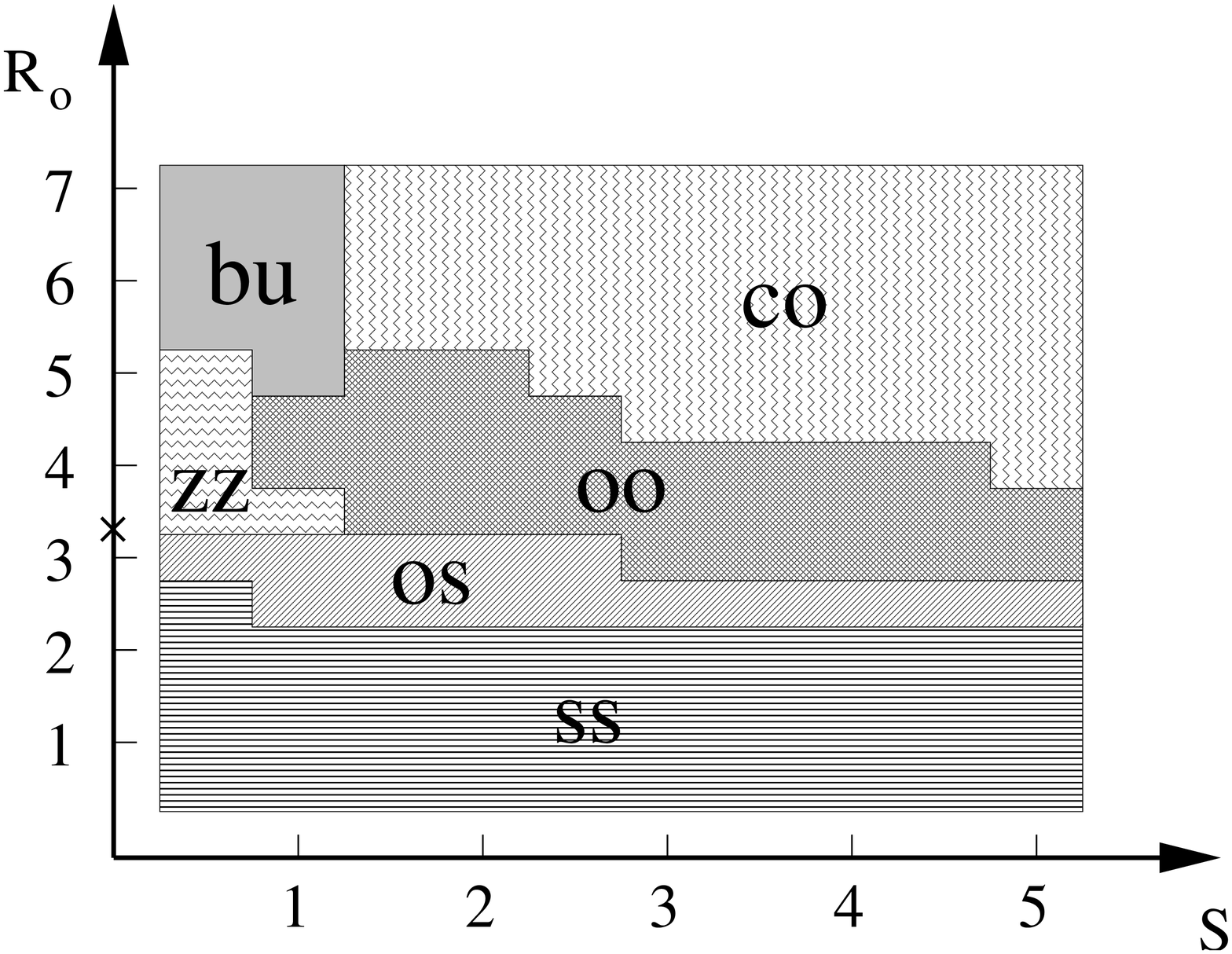}}
\caption{Phase diagram of migration modes in the plane of 
anisotropy strength $S$ and island radius $R_0$, for sixfold anisotropy
($n=6$) and the field aligned with a direction of maximal mobility
($\alpha = 0$). For each point on
a grid of resolution $0.5 \times 0.5$, the evolution of the island was
followed until the asymptotic mode could be identified. We distinguish
between straight stationary (\textbf{ss}), oblique stationary (\textbf{os}), oblique oscillatory 
(\textbf{oo}), zigzag (\textbf{zz}) and complex oscillatory (\textbf{co}) motion. In the region 
\textbf{bu} islands break up. The cross
on the $R_0$-axis indicates the linear instability of the circular
solution in the isotropic case.} \label{Phase}
\end{figure}

We now turn to the effects of crystal anisotropy. 
Throughout this paper only the anisotropy of the adatom mobility $\sigma$
will be taken into account, while the
edge stiffness $\tilde \gamma$ is kept isotropic. This is motivated partly
by the fact that the anisotropy in $\sigma$ is found experimentally
(to the extent that it has been investigated) to much exceed that of 
$\tilde \gamma$ \cite{Giesen04}, and partly by our desire
to disentangle kinetic ($\sigma$) and 
thermodynamic ($\tilde \gamma$) effects \cite{note:stiffness}. 
For the kinetic anisotropy we employ the functional form \cite{Schimschak00}
\begin{equation}
\label{sigma}
\sigma(\theta) = \sigma_{\mathrm{max}} (1 + S)^{-1}  
\{ 1 + S \cos^2 [n(\theta + \alpha)/2] \}.
\end{equation}
Since the prefactor $\sigma_{\mathrm{max}}$ only sets the time scale, the relevant parameters
in (\ref{sigma}) are the anisotropy strength $S$, the number of symmetry axes
$n$, and the angle $\alpha$ between the symmetry axes and the electric
field direction. The natural time unit is
$t_E = l_E^4/(\sigma_{\mathrm{max}} \tilde \gamma \Omega^2)$ \cite{Schimschak00}.

The simplest solutions in the anisotropic case are stationary
islands moving at constant speed, which satisfy the equation
$v_n = V \sin(\theta + \phi)$; here the angle $\phi$ accounts for the fact that
the island does not necessarily move in the field direction. A complete
analysis of stationary island shapes has been achieved
in the limiting case of zero stiffness \cite{Kuhn04}. For an even number
$n$ of symmetry axes smooth stationary
shapes are found for small $S$, while for larger anisotropy the shapes develop
self-intersections; no stationary shapes exist when $n$
is odd. 
The migration direction generally lies between the field
direction and the symmetry axis of the anisotropy. 

Despite their mathematical interest, these results are of limited applicability
to real islands, because all stationary shapes are wildly unstable when
$\tilde \gamma = 0$. In the remainder of the article we therefore focus on
the numerical solution of the full, time-dependent
problem with $\tilde \gamma > 0$ and $\sigma(\theta)$
given by (\ref{sigma}). Two complementary numerical algorithms have been 
employed. For relatively small islands a finite difference scheme 
described in \cite{Schimschak00} was found to be most efficient, while for large islands
we rely on the better 
stability properties of a semi-implicit adaptive finite element algorithm. 
\cite{Baensch05}.
The full mutual consistency of the two approaches has been checked.

Most results have been obtained for $n=6$ and $\alpha = 0$.
This leaves the anisotropy strength $S$ and the initial condition for
the deterministic shape evolution
to be specified. Extensive calculations show that the dependence
on the precise initial shape is minor \cite{note:initial}, 
and hence the initial
condition can be characterized by the radius $R_0$ of a circular
island of the same area; in practice, we usually start the calculation
from a slightly distorted circle. 

\begin{figure}
\centerline{\includegraphics[scale=0.25,angle=-90]{%
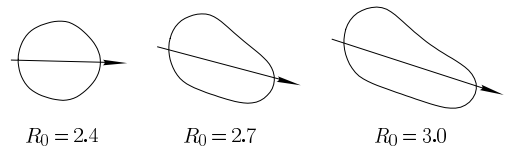}}
\caption{Stationary shapes for $S=2$ near the 
transition from {\bf ss} to {\bf os} behavior. Arrows indicate the direction
of motion.} \label{Oblique}
\end{figure}

The phase diagram in Fig.\ref{Phase} gives an overview of the observed
migration modes in the $S-R_0$-plane \cite{note:other}. 
For small islands ($R_0 \leq 2$)
the evolution converges to a stationary shape which moves in
the direction of the field. For large $S$ the shapes develop
facets \cite{Kuhn04}, similar to what has been observed for void 
electromigration \cite{Schimschak00}. Increasing the island radius
the direction of migration starts to deviate from the field direction,
and we enter the regime of oblique
stationary (\textbf{os}) motion (Fig.\ref{Oblique}). Since the field is aligned with the symmetry
axis of the anisotropy, the appearance of obliquely moving solutions
implies that the symmetry of the problem is \emph{spontaneously}
broken. In the {\bf os} regime, pairs of symmetry-related stationary
solutions coexist; which solution is chosen in a given
run depends on the initial condition.     

\begin{figure}
\centerline{\includegraphics[scale=0.37]{%
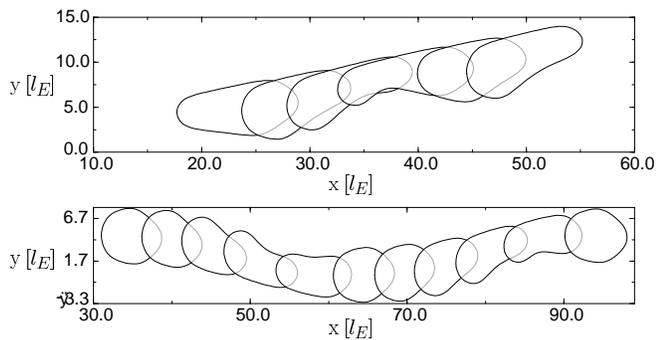}}
\caption{Snapshots of \textbf{oo} motion for
$R_0 = 4$ and $S=1$ (upper panel) and \textbf{zz}  motion for
$R_0 = 3.5$, $S=0.5$ (lower panel), taken at
time intervals $\Delta t = 20$. In both cases
the perimeter displays simple oscillations, 
as in the bottom panel of Fig.\ref{Osci}.} 
\label{Zigzag}
\end{figure}

\begin{figure}
\centerline{\includegraphics[scale=0.35]{%
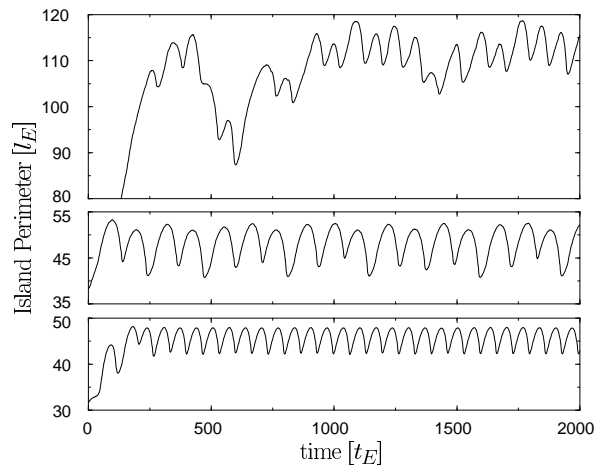}}
\caption{Time series of the island perimeter showing regular
and irregular oscillations. From bottom to top the parameters
are $S=2,R_0=5$; $S=5,R_0=5$; and $S=3,R_0=8$.
The top panel corresponds to the run shown in Fig.\ref{Chaos}.} 
\label{Osci}
\end{figure}

\begin{figure}
\centerline{\includegraphics[scale=0.5,angle=-90]{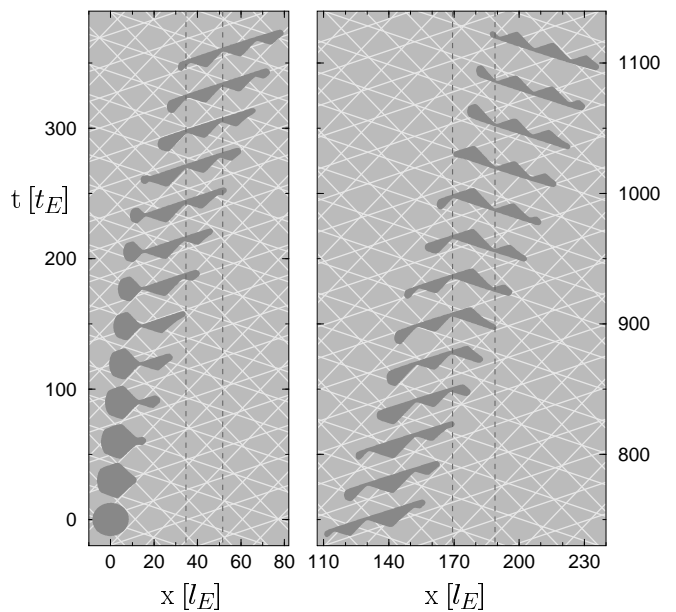}}
\caption{Complex oscillatory motion with 
$S=3$, $R_0 = 8$. Light lines show the 
facet orientations predicted from (\ref{Newton}), dark dashed lines illustrate 
that the hill-valley structure is static in the substrate frame.
Consecutive snapshots are shifted upwards in time.} \label{Chaos}
\end{figure}

Upon further increase of $R_0$ the obliquely moving shapes start
to oscillate (Fig.\ref{Zigzag}). Near the onset
of oblique oscillatory (\textbf{oo}) motion at radius $R_0^c$ 
the oscillation period diverges
as $T \sim \vert R_0 - R_0^c \vert^{-\nu}$ 
with $\nu \approx 2.5$. For larger radii higher harmonics of the
fundamental oscillation period appear and the motion becomes increasingly
irregular (Fig.\ref{Osci}). 
This characterizes the complex oscillatory (\textbf{co}) regime,
which is exemplified in 
Fig.\ref{Chaos}. The direction of island motion displays 
random shifts, which seem to be 
triggered by small fluctuations. 
This behavior is typical for large islands, and it
is distinct from the \textit{periodic} direction changes seen in the
zig-zag (\textbf{zz}) regime for moderate sizes and small anisotropies 
(Fig.\ref{Zigzag}).

The true long time behavior for large islands ($R_0 > 7$) and 
large anisotropy ($S > 5$) could not be pinned down unambiguously with
our current numerical methods. Generally speaking, large islands with small
anisotropy break up, while for large $R_0$ and $S$ 
facetted shapes undergoing irregular motion dominate.

The example shown in Fig.\ref{Chaos} provides an 
important clue to the origin of the complex shape evolution.
After an initial transient lasting until $t \approx 300$, the 
island settles down into a shape consisting of 
a straight upper edge and
a lower edge which has broken up into a facetted hill-and-valley
structure. The direction of island motion coincides
with the orientation of the upper, straight edge, as shown for 
a smaller island in the upper panel of Fig.\ref{Zigzag}.
The key observation is that the
hill-and-valley structure on the facetted edge \emph{does not move} 
in the substrate frame. The moving island slides past the static
facets, causing the shape to oscillate. Around $t \approx 900$ the roles of 
the upper and lower edges are seen to reverse, and the direction of motion changes. 

Quite generally, large islands in the \textbf{co} regime can be constructed from
four selected facet orientations $\theta_1, \theta_2, \theta_3, \theta_4$. Here
$\theta_1$ and $\theta_2$ are the possible stable orientations of the upper
island edge, which must satisfy 
$
- \pi/2 < \theta_1 < 0 < \theta_2 < \pi/2.
$
In the case considered here ($\alpha=0$, $n$ even)
the corresponding orientations $\theta_3$ and $\theta_4$ for the lower
edge are obtained by reflection at the $x$-axis,  
$\theta_3 = - \pi - \theta_1$ and $\theta_4 = \pi - \theta_2$. 
To form a closed shape, at least three orientations
must be combined, two of which are those two symmetry-related
orientations that are closest to the horizontal direction ($\theta = 0$
or $\pi$). In Fig.\ref{Chaos}
we see a transition from a shape with orientations 
$\{\theta_1,\theta_3,\theta_4\}$ to a shape with orientations
$\{\theta_1,\theta_2,\theta_3 \}$.

The stable facet orientations can be computed along the lines of 
\cite{Krug94}. The condition $v_n = 0$ for a static shape implies that the
current in (\ref{cont}) is set to a constant
$j^\ast$. Using the relation $\kappa = d \theta/ds$ this can be 
brought into the form
\begin{equation}
\label{Newton}
\Omega \tilde \gamma \frac{d^2}{ds^2} \theta = - [j^\ast/\sigma(\theta) - q^\ast E_0 \cos(\theta)]
\equiv -V'(\theta),
\end{equation}
which describes the position $\theta(s)$ of a particle moving in time $s$ subject
to a potential $V(\theta)$ determined by the mobility and the electric
field strength. As explained in \cite{Krug94}, the coexistence of two stable
facet orientations corresponds to a particle trajectory moving between two
degenerate potential maxima. 
To determine the selected orientations, $j^\ast$ is tuned until two 
degenerate maxima satisfying the above constraints appear. 
We have checked that this procedure correctly accounts
for the facet orientations observed in the time-dependent calculations throughout
the relevant region of the phase diagram (see Fig.\ref{Chaos}). 
In general, stable facets can
be constructed from (\ref{Newton}) only when the anisotropy is sufficiently large \cite{Krug94}.
For $n=6$, $\alpha=0$ the requirement is $S > S_c \approx 1.77$.
No stable facets are found when the number of symmetry axes is too small 
($n \leq 3$); this may explain why we do not see 
oscillatory shape evolution for a threefold anisotropy. 


We close with two remarks concerning future research. First,  
we note that
the observed island shapes are quite smooth, which implies that the number
of circular harmonics involved in the shape evolution is small. It thus seems
promising to attempt a description in terms of a low-dimensional dynamical
system, in the spirit of \cite{Valance92}, to gain a deeper understanding
of the various migration modes and the bifurcations connecting them. 
  
Second, we address the experimental conditions under
which the predictions of this paper could be realized. 
As an example, we consider islands on Cu(100), for which most
material parameters entering the theory are available. Following
\cite{Mehl00}, we estimate that the electromigration force on an
edge atom at a current density of 
$10^7 \textrm{A} \textrm{cm}^{-2}$ is about 400 eV/cm. Together with
the experimentally determined stiffness \cite{Dieluweit03} and mobility
\cite{Giesen04} for kinked steps at 300 K,
this yields a characteristic length of $l_E \approx 25 \; \textrm{nm}$,
and a time scale $t_E$ on the order of seconds.
Thus we expect complex shape dynamics to be observable for island 
radii around $100 \; \textrm{nm}$ and on time scales of a few hundred
seconds. As a first step
towards a more detailed description of specific surfaces, it would be 
important to identify oscillatory shape evolution in kinetic Monte Carlo
simulations of island electromigration.

\begin{acknowledgments}

This work was supported by DFG within SFB 616 
\textit{Energy dissipation at surfaces} and SFB 611 \textit{Singular phenomena
and scaling in mathematical models.} 

\end{acknowledgments}


\begin{thebibliography}{99}
\bibitem{note:initial} In some cases, however, an elongated initial shape can be stabilized
for island sizes where a more circular initial condition leads to breakup.
\bibitem{note:other} Similar migration modes have been observed for $n=4$ and for 
$n=6$ with other values of $\alpha$, with the expection of the case $n=6$ and
$\alpha = \pi/6$, where only straight moving stationary shapes are observed.
\end{thebibliography}
\end{document}